\documentstyle[11pt,aaspp4]{article}
\tighten
 
\received{31 Dec 1999}

\slugcomment{accepted by MNRAS}

\begin{document}

\title{Cosmic Momentum Field and Mass Fluctuation Power Spectrum}
\author{Changbom Park }
\affil{Department of Astronomy, Seoul National University,
    Seoul, 151-742 Korea}
\affil{cbp@astro.snu.ac.kr}

\begin{abstract}
We introduce the cosmic momentum field as a new measure of
the large-scale peculiar velocity and matter fluctuation fields.
The momentum field is defined as the peculiar velocity field traced and
weighted by galaxies, and is equal to the velocity field in the linear
regime. 
We show that the radial component of the momentum field can be considered 
as a scalar field with the power spectrum which is practically 
$1/3$ of that of the total momentum field.  
We present a formula for the power spectrum
directly calculable from the observed radial peculiar velocity data.

The momentum power spectrum is measured for the MAT sample in the 
Mark III catalog of peculiar velocities of galaxies. 
Using the momentum power spectrum
we find the amplitude of the matter power spectrum is
$6400^{+2800}_{-1800}$ and $4500^{+2000}_{-1300}$ $\Omega^{-1.2} 
\;(h^{-1} {\rm Mpc})^{-3}$
at the wavenumbers 0.049 and 0.074 $h$ Mpc$^{-1}$, respectively, where 
$\Omega$ is the density parameter.
The 68\% confidence limits include the cosmic variance.
The measured momentum and density power spectra together indicate 
that the parameter $\beta_O = \Omega^{0.6}/b_O
= 0.51^{+0.13}_{-0.08}$ or 
$\Omega = 0.33^{+0.15}_{-0.09} \; b_O^{5/3}$ where $b_O$ is the bias
factor for optical galaxies.

\end{abstract}


\section{INTRODUCTION}



There are advantages and disadvantages in using the absolute distance
and peculiar velocity data to explore the large scale structure.
The large scale structures revealed by these data are in the real space, 
and are free of the redshift distortions. Furthermore,
unlike the galaxy distribution studies, the peculiar velocity 
of galaxies gives us direct information on the mass fluctuation field.
Since both galaxy and matter fluctuation fields can be obtained from 
such data, we can also compare them to each other to study 
the biasing of galaxy distribution.

A major disadvantage of using the peculiar velocity data is 
that the positions of galaxies and thus their peculiar velocities
have large errors proportional to distance. 
Because measuring the absolute distance is harder compared
to the redshift measurement, the samples in general are much smaller 
than redshift survey samples. This
disadvantage is partly relieved by the fact that the observed velocity
field contains information on larger scales compared to the density field
in a given survey volume. 
And the unique opportunity to directly probe the large scale
 matter density field 
makes the peculiar velocity observation very important
in the study of large scale structure.

In the past decade or so there have been major progresses in the study 
of the velocity field, both observationally and theoretically. 
Observationally, many surveys of galaxy peculiar velocities
have been made. Distances to galaxies are measured by using the 
Tully-Fisher or the Dn-$\sigma$  methods for spirals and ellipticals, 
respectively.  
An important data set in the recent years has been the Mark III
catalog (Willick  et al. 1997). It currently contains about 3,300 
galaxies whose distances are measured upto $17 \sim 21$\%.
More recently completed survey data are the SCI catalog containing 
782 galaxies in 24 clusters of galaxies, and 
the SFI catalog 
of 1216 Sbc - Sc galaxies (da Costa et al. 1996; Giovanelli et al. 1997).
On-going surveys include the ENEARf sample of about 1400 field 
early type galaxies, the ENEARc of about 500 ellipticals and S0's
in the nearby 28 clusters (Wegner et al. 1999), 
the EFAR of 736 early-types in 85 clusters (Colless et al. 1999), 
the SMAC of 699 early-types in 56 Abell clusters (Hudson et al. 1999),
the SHELLFLOW of 276 Sb - Sc galaxies (Courteau et al. 2000),
and the LP10K of 244 TF galaxies in 15 Abell clusters (Willick 1999),
among others.
The number of galaxies with measured absolute distances is now of the order of
$10^4$. The number of observed galaxy redshifts was approaching that order
in the late 1980's when the study of the large scale 
distribution of galaxies in the redshift space has received a lot 
of attention. 
 
 On the theory side, many analysis methods have been developed and 
applied to the observational data. 
A major development has been the POTENT method
(Bertschinger et al. 1990; Dekel et al. 1999). 
It aims to recover the smoothed three-dimensional peculiar velocity 
field from the observed radial velocity data. 
Other velocity field studies have used the 
velocity correlation function (CF) statistic (Gorski et al. 1989; 
Borgani et al. 2000), the maximum likelihood method
(Freudling et al. 1999), Wiener filtering method (Zaroubi et al. 1999),
the orthogonal mode-expansion method (da Costa et al. 1998;
Nusser \& Davis 1995),
and the $\nabla\cdot {\bf v}$ field method (Bernardeau \&
van de Weygaert 1996). 

From the analysis of the peculiar velocity data one measures the amplitude 
of mass fluctuation and/or the $\beta$ parameter.
Recent studies have shown that different analysis methods can yield
contradicting results even when they are applied to a common data,
and thus the validity of some of these methods is questioned
(Kolatt \& Dekel 1997; Freudling et al. 1999; Borgani et al. 2000;
see Discussion section for more examples and references).
We note that the peculiar velocity can be measured only where 
galaxies are observed and that
the physical quantity directly measurable from observed data 
is the momentum field rather than the velocity field. 
If one deals with the momentum field, there is no need to fill up 
the voids with interpolated velocities by a large smoothing 
since the voids do not carry momentum by definition. 
We also adopt to use the power spectrum (PS) instead of CF to
localize the effects of noise in observational data at small scales.
Furthermore, the PS of the momentum field is observationally calculable and
easily compared with theories. It can be estimated 
directly from the observed radial velocity data at 
galaxy positions just as the power spectrum of the galaxy 
density field can be estimated directly from galaxy positions 
(cf. Park et al. 1994).
The PS in the linear regime, which is equal to that of the linear velocity 
field, can be directly compared with cosmological models.



In the following section we define the momentum field 
as the peculiar velocity field traced and
weighted by galaxies. It is equal to the velocity field in the linear
regime. The PS of the momentum field is measured from an observed 
peculiar velocity data and used to
estimate the amplitude of the total mass fluctuation.

\section{Theory}

\subsection{Definitions}

The physical observable we use is the radial components of the peculiar 
velocities sampled at locations of galaxies. In this section I present 
the theory necessary to link this observational
quantity to cosmological models.
Suppose the matter density $\rho({\bf x})$ and the peculiar velocity
${\bf v}({\bf x})$ fields are homogeneous and isotropic random fields. 
The density and velocity fields can be non-linear (section 2.4), and
galaxies can be biased tracers of matter (section 2.5).
Instead of the density we will use the normalized density 
$\rho /{\bar \rho}=1+\delta({\bf x})$, where $\delta$ is the overdensity
field and ${\bar \rho}$ is the mean density. We define the momentum field as 
$${\bf p} = (1+\delta) {\bf v}, \eqno (1)$$
which has the dimension of velocity and is equal to ${\bf v}$
in the linear regime.
Its radial component $p_r = (1+\delta) v_r$ is directly observable
from a peculiar velocity survey.

The Fourier transforms (FTs) of these fields are defined in a large
volume $V$ over which they are considered to be periodic. For example,
the FT of the momentum field is defined as 
$${\bf p}_{\bf k}
= \int {{d^3 x}\over V} {\bf p}({\bf x})
  e^{i{\bf k}\cdot{\bf x}}. \eqno (2)$$

\subsection{The correlation function and power spectrum}

To compare the observed peculiar velocities of galaxies
with cosmological models we measure the PS
or equivalently the CF of the momentum field.
The general form of the two-point correlation tensor of an
isotropic momentum field is
$${\xi}^{p}_{ij} = \langle p_i ({\bf x}) p_j ({\bf x}+{\bf r})\rangle
    = [{\xi}^p_{LL}(r)-{\xi}^p_{NN}(r)]{\hat r}_i {\hat r}_j 
      +{\xi}^p_{NN} (r) \delta_{ij}, \eqno (3)$$
where ${\hat r}_i = r_i /r$, and ${\xi}^p_{LL}$ and ${\xi}^p_{NN}$ are 
the CFs of the momentum field 
in the directions parallel and perpendicular to the separation vector
${\bf r}$ between two points (Monin \& Yaglom 1971, p39; hereafter MY). 
It can be shown that the dot-product 
CF of the momentum field is (MY; Peebles 1987;
Gorski 1988; Kaiser 1989; Szalay 1989; Groth et al. 1989)
$${\xi}_p(r)= {\xi}^p_{ii} = 
       \langle {\bf p} ({\bf x})\cdot {\bf p}({\bf x}+{\bf r})
        \rangle = {\xi}^p_{LL}+2 {\xi}^p_{NN}. \eqno (4)$$

The spectral tensor $P^p_{ij}(k)$ of the momentum field can be defined 
as the FT of the correlation tensor. In particular, the trace 
$P_p(k)=P^p_{ii}(k)$ is the FT of the dot-product CF
$$P_p(k) = 
         \int {d^3x \over V} {\xi}_p(r)e^{i {\bf k}\cdot{\bf r}},
         \eqno (5)$$
which can be shown by using equations (2) and (4).
In the linear regime the continuity equation yields
$P_p(k) = \langle |{\bf p}_{\bf k}|^2\rangle
=\langle |{\bf v}_{\bf k}|^2\rangle = (DHf/k)^2 P_{\delta}(k)$ where
$D$ is the linear growth factor as in $[\delta({\bf x},t)]_{\bf k} =
D(t)\delta_{\bf k}$, $H$ is the Hubble parameter,
$f=a{\dot D}/{\dot a}D\approx\Omega^{0.6}$, and $a$ and $\Omega$ are the expansion 
and density parameters.
One can estimate the linear matter
 density PS from the linear part of the measured momentum PS, i.e.
$P_{\delta}(k)= (k/D H f)^2 P_p (k)$.

\subsection{Radial component of the momentum field}

We hope to compare the observed peculiar velocity field with
predictions of various cosmological models. Unfortunately, 
only the radial components of the three-dimensional velocity vectors 
can be observed, and furthermore they can be observed only
at positions of galaxies. 
The data has often been weighted by volume to give the radial component 
of the `continuous' peculiar velocity field. However,
galaxies as the velocity tracers are missing in voids
where interpolations are needed, and are
numerous at clusters where the volume weighting makes one lose 
information in the data. Reconstruction of the velocity field from 
volume-weighting of the observed peculiar
velocity data thus requires a large smoothing which leaves us a very small
usable dynamic range in the data (Dekel et al. 1999).
What we directly measure is the galaxy number-weighted quantity
$p_r = v_r \rho_g /{\bar \rho_g}$. 
Here the radial peculiar velocity $v_r$ is caused by the total matter field 
and $\rho_g /{\bar \rho_g}$ represents the distribution of galaxies which
are the tracers of both mass and velocity.


We now find the statistical relation of the field $p_r$ with 
the total momentum field ${\bf p}$. 
Consider two galaxies at positions ${\bf r}_1$ and ${\bf r}_2$ separated
by the vector ${\bf r}$ which subtends $\theta$ in angle on the sky.
Let the angles between ${\bf r}$ and ${\bf r}_i$ are $\gamma_i$ so that
$\theta = \gamma_1 - \gamma_2$. Then, the CF of the radial component
of the momentum field is from equation (3)
$$\langle {\bf p}({\bf r}_1)\cdot {\hat{\bf r}}_1\;
 {\bf p}({\bf r}_2)\cdot {\hat{\bf r}}_2\rangle
=({\xi}^p_{LL}-{\xi}^p_{NN}){\hat{\bf r}}_1\cdot {\hat {\bf r}}\;
                {\hat{\bf r}}_2\cdot {\hat {\bf r}}
    +{\hat{\bf r}}_1\cdot {\hat{\bf r}}_2 {\xi}^p_{NN}$$
$$ =\cos{\gamma}_1 \cos{\gamma}_2 \; {\xi}^p_{LL}(r) 
  +\sin{\gamma}_1 \sin{\gamma}_2 \; {\xi}^p_{NN}(r). \eqno (6)$$
Therefore, the radial component of the momentum field does not depend
just on $r$, but also on the angle variables $\gamma_1$ and $\gamma_2$.
However, we can find a very useful formula by averaging it out 
over these angles. Suppose the angle $\theta=\gamma_1 -\gamma_2
\ll 1$ so that ${\bf r}_1$ and ${\bf r}_2$ are approximately parallel. 
Then $\langle p_r({\bf r}_1) p_r({\bf r}_2)
\rangle\approx \cos^2 \gamma_1 {\xi}^p_{LL} +
\sin^2\gamma_1 {\xi}^p_{NN}$. The vector ${\hat{\bf r}}_1$ covers the surface
of a unit sphere unless the survey boundaries prohibit. If the
survey boundary effects are negligible, $\langle \sin^2 \gamma_1\rangle
=2/3 = 2\langle \cos^2\gamma_1\rangle$. Equation (6) and (4) then implies
$${\xi}_{p_r} (r) = \langle p_r ({\bf x}) p_r ({\bf x}+{\bf r})\rangle
  \approx {1\over 3} {\xi}_p(r). \eqno (7)$$
Therefore, at separation scales small compared to the distance from
the origin we can treat the radial component of the momentum field as
an isotropic scalar field. And its CF or PS measured from observations
can be directly compared with theories.
Equation (6) has been also derived by Kaiser (1989), Szalay (1989), and
Groth et al. (1989), who have tried to estimate ${\xi}^p_{LL}$
and ${\xi}^p_{NN}$ in the $(r, \gamma_1, \gamma_2)$ space.
The relation (7) has been empirically noted by Gorski et al. (1989),
but only to claim its inferiority against their statistics.

The PS of this radial momentum scalar field is defined as the FT of 
the CF ${\xi}_{p_r} (r)$
$$P_{p_r} (k) = \int  {{d^3 x}\over V} {\xi}_{p_r} (r) 
       e^{i{\bf k}\cdot{\bf x}}\approx {1\over 3} P_p(k). \eqno (8)$$
Equation (8) has been derived only by assuming the isotropy
of the momentum field and with the far-field approximation, 
and should also hold in the non-linear regime.
Using equation (8), one can compare
the observed radial momentum field PS with the linear theory
or results of simulations of cosmological models.
We demonstrate in Figure 1 that equation (8) is actually 
very accurate over wide scales when the cosmic variance and 
observational uncertainties in the PS are taken into account.
We have generated a set of linear momentum fields of the open CDM
model 
in a 512 $h^{-1}$ Mpc box
with $\Omega h=0.2$ with the normalization of $\sigma_8=1$,
the RMS fluctuation of density within an 8 $h^{-1}$ Mpc sphere.
The PS of the total momentum (filled dots),
or velocity in this linear experiment,
is shown to be nearly the same as three times those of the radial momentum
observed at a corner (squares) or at the center (triangles)
of the simulation cube. The worst case
is for the fundamental mode and when the observation is over all sky
(The observer is at the center). Even in this case
the PS of the radial momentum is only 14\% lower than that of 
the total momentum.
From now on, we consider $ 3P_{p_r} (k)$ as $P_p(k)$.
We have also noted that the statistical fluctuation of $P_{p_r} (k)$
is smaller than the fluctuations 
of the PS of the individual $x, y, z$ components
of the momentum vector. 

It is easy to calculate the PS of the radial momentum field from the observed
peculiar velocity data. Suppose we have 
radial components of peculiar velocities of $N$ galaxies. The FT of
the observed radial momentum field is
$$p_{r{\bf k}} = \int {{d^3 x}\over V}
      ({\rho\over{\bar \rho}}{\bf v}\cdot{\hat{\bf x}})
        e^{i{\bf k}\cdot{\bf x}}=
\sum_{j=1}^{N} v_r ({\bf x}_j)  e^{i{\bf k}\cdot{\bf x}_j}, \eqno (9)$$
where it is assumed that galaxies represent mass, and
$\rho({\bf x})/{\bar \rho} = \sum_j \delta^D ({\bf x}-{\bf x}_j )$.
The PS $P_{p_r} (k)$ of the radial momentum field can be estimated by 
simply taking averages of $|p_{r{\bf k}}|^2$.
Without the factor $v_r ({\bf x}_j)$ equation (9) yields the FT of
the density field. 
Real surveys do not cover
the entire universe, and the data can have different statistical 
weights for different galaxies due to the variation of selection. 
Modification of the formula can be made in a way similar 
to the case of the galaxy density field (Park et al.  1994).

The final formula for the estimated PS of the momentum field is
$$ P_p(k) = 3\left[ \langle|{\hat p}_{r{\bf k}}|^2\rangle
     -{\sum w^2_j\over (\sum w_j)^2}\right] /\sum_{\bf k} |W_{\bf k}|^2, \eqno (10)$$
where
$$ {\hat p}_{r{\bf k}}=\sum_j w_j v_{rj} e^{i{\bf k}\cdot {\bf x}_j}
  /\sum w_j -W_{\bf k}, \eqno (11)$$
$w_j$ are weights to galaxies, and $W_{\bf k}$ are the FT of the survey
window function (see sections 3.1 and 3.3).
Corrections for the effects of errors in peculiar velocities and 
distances are made by using Monte Calro simulations explained in section
3.3.

\subsection{Difference between the momentum and velocity fields}

The momentum field ${\bf p}=(1+\delta){\bf v}$
differs from the velocity field ${\bf v}$ by $\delta{\bf v}$.
It is instructive to understand the nature of the newly added field
$\delta {\bf v}$, which dominates the momentum field in the non-linear 
regime. Since this field is a multiplication of $\delta$ to ${\bf v}$,
its FT is the convolution between the fields $\delta_{\bf k}$ and
${\bf v}_{\bf k}$
$$ [\delta ({\bf x}) {\bf v}({\bf x})]_{\bf k} = 
   \int {{d^3 k'}\over (2\pi)^3} \delta_{{\bf k}'}
        {\bf v}_{{\bf k}-{\bf k}'}. \eqno (12)$$
To see the behaviour of this field specifically, we
substitute the linear velocity ${\bf v}_{\bf k} = iDHf{\bf k}
\delta_{\bf k}/k^2 $ in equation (12) to obtain
$$ [\delta ({\bf x}) {\bf v}({\bf x})]_{\bf k} = 
  iD^2 H f  \int {{d^3 k'}\over (2\pi)^3}{{\bf k}'\over k'^2}
  \delta_{{\bf k}'}\delta_{{\bf k}-{\bf k}'}. \eqno (13)$$
We find that the PS of the $\delta {\bf v}$ field is
$$\langle |(\delta {\bf v})_{\bf k} |^2 \rangle
  = {1\over 2} (D^2 Hf)^2  \int {{d^3 k'}\over (2\pi)^3}
   [{{\bf k}'\over k'^2}+{{{\bf k}-{\bf k}'}\over {|{\bf k}-{\bf k}'|^2}}]^2 
    P(k') P(|{\bf k}-{\bf k}'|), \eqno (14)$$
where we have used the formula for a Gaussian field $\delta_{\bf k}$
$$\langle \delta_{{\bf k}'}\delta_{{\bf k}-{\bf k}'}
  \delta_{-{\bf k}''}\delta_{-{\bf k}+{\bf k}''}\rangle$$
$$
  =P(k')P(k'')\delta_D^2({\bf k})+
   P(k')P({\bf k}-{\bf k}')
      \delta_D({\bf k}'-{\bf k}'')\delta_D({\bf k}'-{\bf k}'')+$$
$$
   P(k')P({\bf k}-{\bf k}')
      \delta_D({\bf k}'-{\bf k}+{\bf k}'')\delta_D({\bf k}-{\bf k}'-{\bf k}'').
  \eqno (15)$$
Here $\delta_D$ is the Kronecker delta.
This PS can be numerically integrated for a given density PS.
For $P(k)\propto k^n$ 
it can be analytically shown that 
$\langle |(\delta {\bf v})_{\bf k} |^2 \rangle \propto k^n$
if $n < -1$, and $\propto k^2$ if $n > 1/2$. In the case of the Standard
CDM (SCDM) model the slope of the PS of the $\delta {\bf v}$ field
changes from $-3$ at small scales to $+2$ at large scales.
Figure 2 shows the linear density, and velocity power spectra $P_{\delta}$
and $P_v$ of the SCDM model, and the corresponding 
$P_{\delta{\bf v}}$ calculated from equation (14).
Even though the observable quantity is $(1+\delta) {\bf v}$
and not $\delta {\bf v}$, Figure 2 reveals an important property of
the $\delta {\bf v}$ field. That is, its PS has a peak at a 
scale much smaller than that of the density PS. 
The SCDM density PS has a peak at $k\simeq 0.05 h$ Mpc$^{-1}$
or $\lambda \simeq 125 h^{-1}$ Mpc. But the PS of $\delta {\bf v}$ has a peak
at $k\simeq 0.158 h$ Mpc$^{-1}$ or $\lambda \simeq 40 h^{-1}$ Mpc. 
Furthermore, the width of the peak is much narrower.
One can, in principle, obtain this field  $\delta {\bf v}
={\bf p}-{\bf v}$ by subtracting the volume-weighted velocity field
from the mass-weighted momentum field both calculated from
observed peculiar velocity data.

N-body simulations show that in the strongly non-linear regime 
the density PS approaches to the slope $n=-1$ for CDM-like
hierachical models. 
It is expected that the PS of the momentum field
will then have the corresponding slope of about $-0.8$
(calculated from equation 14) at small non-linear
scales where the momentum field is dominated by the $\delta {\bf v}$ field.
On the other hand, at very large linear scales the PS of the momentum field  
will be equal to that of the velocity field. In universes with the
Harrison-Zel'dovich primordial density PS with $n=1$ at large scales, the
corresponding slope of the momentum field PS is $n-2 = -1$.
Therefore, the slope of the momentum field PS should vary from $-1$
at very large scales, to $-2$ near the peak of the density PS,
and then back to about $-1$ at small scales.
The deviation of the PS from the linear one and its transition 
to the characteristic small scale slope marks the non-linear scale.

On the other hand, the non-linear part of the velocity PS 
decreases very rapidly and is hard to measure from an observed data.
The non-linear part of the momentum PS, decreasing 
only as $\sim k^{-1}$, is easier to measure and useful
for the model discrimination.

\subsection{Case of the biased galaxy distribution}

In the discussions above we have assumed that the galaxy distribution
represents the mass field. If this is not true, the formulae containing
the non-linear term $\delta {\bf v}$ should be modified because
the observed quantity is now $(1+\delta_g) {\bf v}$.
Suppose galaxies are linearly biased with respect to 
mass by a constant, that is, ${\delta_g}_{\bf k}=b\delta_{\bf k}$ 
where $b$ is the bias factor. 

In the linear regime the PS calculated from equation (10)
approximates to
$$P_p(k) = \langle|{\bf v}_{\bf k}|^2\rangle
    = b^{-2} (DHf)^2 k^{-2} P_{\delta_g}(k).
      \eqno (16)$$
In the non-linear regime the PS depends also on two-point functions of
the third and fourth orders of $\delta$ and ${\bf v}$.
The fourth order term will be dominant in the non-linear regime.
If equation (14) is used, it becomes
$$P_p (k) \approx b^2 \langle |(\delta {\bf v})_{\bf k} |^2 \rangle$$
$$ \approx {1\over 2}b^{-2} (D^2 Hf)^2  \int {{d^3 k'}\over (2\pi)^3}
   [{{\bf k}'\over k'^2}+{{{\bf k}-{\bf k}'}\over {|{\bf k}-{\bf k}'|^2}}]^2
    P_{\delta_g} (k') P_{\delta_g} (|{\bf k}-{\bf k}'|). 
     \eqno (17)$$
Therefore, both linear and non-linear parts of the momentum field PS 
calculated from equations (10) and (11)
depend on $b$ in the same way.
If $b$ is independent of scale, only the overall amplitude of the PS will
be scaled by the combined factor  $ f^2(\Omega)/b^2$
by the biasing, and its shape remains the same.  
The galaxy density PS $P_{\delta_g}$ can also be measured
self-consistently from the same data
using equations (10) and (11) with $v_r$ replaced by 1. 
Then the parameter $\beta= f/b$ can be estimated from
$\beta = (k/DH) (P_p / P_{\delta_g})^{1/2}$ at linear scales.

\section{Experiments in Simulated Universes}

\subsection{Recovering completeness and homogeneity}

Before we apply our method to observed peculiar velocity data,
we make experiments on mock surveys in simulated universes.
From these experiments we can learn about how accurately our method
measures the momentum PS compared to the true one when
the actual data is incomplete, inhomogeneous, and biased.

We first take into account the fact that the real survey
does not cover the whole sky in angle and depth. The incompleteness of
the observed galaxy density and momentum distributions is described
by the survey window function $W({\bf x})$ which is 1 inside
the survey region and 0 outside.
Contribution of the survey window to the estimate of the PS is eliminated
as in equation (11).

An actual survey is usually magnitude or diameter-limited. This means that
the number density of tracers decreases radially
outwards. A best way to remove this
gradient is to make volume-limited samples by using an absolute magnitude
or diameter cut so that the selection criterion on galaxies be
spatially uniform. 
However, when samples are not large, making volume-limited samples
demands too much sacrifice and the resulting subsamples can be too 
small to yield measures of statistics with reasonable accuracies.
We, therefore, use full data of magnitude or diameter-limited surveys 
and correct them for the radial gradient by weighting galaxies with 
$w_j = \phi^{-1}(r_{j})$,
the inverse of the selection function which is the mean number density 
of galaxies at $r_j$ for the given survey selection criteria.
If the selection of the survey in angle (on the sky) is not uniform
due to the Galactic obscuration, for example,
it can be also corrected by using weights.

The accuracy of the galaxy distribution in redshift space is 
almost uniform from nearby to distant regions because
redshifts of galaxies are very accurately measured. 
However, the error in the absolute distances and 
the peculiar velocities of galaxies
monotonically increases radially outward, and will dominate the signal
beyond a certain distance.
To reduce the effects of the noisy data we weigh each galaxy by 
$w_j = 1/(1+\sigma^2_j/\sigma^2_g)$ where $\sigma_j$ is the RMS peculiar
velocity error at distance $r_j$, and $\sigma_g$ is the RMS 
one-dimensional velocity dispersion of galaxies. The weight is nearly
constant until $\sigma_j \sim \sigma_g$, and starts to drop rapidly.
Inclusion of these weights in the theory
is made as in equations (10) and (11) with all above weights 
multiplied together (see also Park et al. 1994).

\subsection{Correction for Malmquist bias}

The most serious trouble in analyzing peculiar velocity
data is making corrections for the Malmquist bias (Lynden-Bell et al. 1988). 
Malmquist biases are caused by the random error in the galaxy distances 
estimated by the distance estimators like Tully-Fisher (TF) or 
$D_n- \sigma$ methods (cf. Willick et al. 1996), which
monotonically increases as the distance increases.

Biases in the inferred distance, and therefore in the peculiar velocity
occur due to the volume effect and galaxy number density 
variations along the line-of-sight. 
That is, at a given distance there are in general 
more galaxies perturbed from the far side than from the near side.
The relative number of galaxies randomly moved from inside and outside 
depends also on the actual distribution of galaxies. This combined effect
is called the inhomogeneous Malmquist bias (IHM). In the Mark III catalog 
a model of galaxy number density distribution 
$n(r)$ derived from the IRAS 1.2Jy redshift survey (Fisher et al.
1995) has been adopted assuming $\beta=0.6$ and the linear correction
for the peculiar velocity effects.

To imitate the distance error in the mock survey data drawn from
simulated universe models we randomly perturb the true distance
in accordance with the TF magnitude scatter $\sigma$.
After degrading the data we then correct the sample 
for the IHM by estimating the `true'
distance from the formula (Willick et al. 1997)
$$E(r|d) = \int_0^{\infty} r^3 n(r) \exp[-(\ln r/d)^2/2\Delta^2] dr
 /\int_0^{\infty} r^2 n(r) \exp[-(\ln r/d)^2/2\Delta^2] dr, \eqno (18) $$
where $d$ is the given/observed perturbed distance, 
$\Delta=(\ln 10/5)\sigma$,
and $n(r)$ is obtained from the `galaxy' density smoothed 
over 5 $h^{-1}$ Mpc Gaussian in the simulation.

Real data are often selected by magnitude or diameter limits, and/or
by redshift limits, which also causes bias
in the calibration of the forward TF relation.
The effects of these limits on simulated data
are taken account by using the sample selection function and 
the redshift limit when the IHM is corrected for.

The IHM is expected to be alleviated for galaxies in groups
because the distance error can be reduced by a factor of $N^{-1/2}$
for a group with $N$ observed galaxies. 
The redshift-space and TF-distance space proximity conditions adopted by
Willick et al. (1996) and Kolatt et al. (1996) are
used to find galaxies in groups. Many `galaxies'
are erroneously grouped by this method, and the systematic
effects of IHM on the momentum PS decrease only a little 
when the grouping method is applied. 
Therefore, in analyzing the simulated survey data we do not attempt to
find groups and assign the average distances to members.

\subsection{Momentum power spectrum from mock surveys}

To prove that the momentum PS is a measurable and useful statistic, we 
use simulated universe models for which the true momentum power spectra are known.
We make mock surveys in these simulations, correct them  
for IHM, and measure the momentum PS to find out the difference
with the true PS and the amount of uncertainty at each mode.
We have made five N-body simulations of cosmological models: (1) the
Standard CDM model with $\Omega=1$, $h=0.5$, and $b=1$ (SCDM); (2) the open
CDM models with $\Omega=0.4$, $h=0.5$, and $b=1$ (OCDM20$a$) and (3)
$b=1.4$ (OCDM20$b$); (4) the open CDM model with $\Omega=0.25$, $h=0.8$, 
and $b=1$ (OCDM20$c$); (5)
the open CDM model with $\Omega=0.3$, $h=0.5$, 
and $b=1$ (OCDM15).  All OCDM20 models have $\Omega h=0.20$.
Simulations are normalized so that $\sigma_{8} = 1/b$ 
at redshift 0. 

A Particle-Mesh code (Park 1990, 1997) 
is used to evolve $256^3$ CDM particles from $z=17$ to 0
on a $512^3$ mesh whose physical size is 512 $h^{-1}$ Mpc.
For the OCDM20$b$ model the `galaxies' as the biased tracers of
the mass are found from the dark matter particles in accordance
with the peak-background scheme with the galaxy peak scale
$R_s=0.71 h^{-1}$ Mpc, the peak height $\nu_{\rm th} = 0.8$,
and the background smoothing scale $R_b = 3 h^{-1}$ Mpc 
(Bardeen et al. 1986; Park 1991). 

Figure 3 shows the momentum and density power spectra calculated
from these simulations using all particles and without distance
errors (symbols). Curves are the linear power spectra. 
The linear density power spectra 
of the OCDM20$b$ and OCDM20$c$ are not drawn because they are equal 
to that of OCDM20$a$ (long-dashed curve).
The $P_p (k)$ of the OCDM20$b$ (dotted) and OCDM20$c$ (dot-dashed) 
models are lower 
in amplitude than that of the OCDM20$a$ by factors of $\beta^2 = 
\Omega^{1.2}/b^2 =$ 0.51 and 0.57, respectively.
It should be noted that
the momentum PS remains linear at scales $k\le 0.07 h$ Mpc$^{-1}$
while they are certainly not linear at $k\ge 0.1 h$ Mpc$^{-1}$
in all models.

Another important fact is that shapes of the momentum power spectra 
are nearly the same in the quasi-linear and non-linear scales
despite of different cosmological and dynamical parameters.
The deviation in the case of OCDM20$b$ at $k> 0.2 h$ Mpc$^{-1}$
is only due to the background smoothing over $3 h^{-1}$ Mpc used 
in the biasing scheme.
This universality of the shape of the momentum PS makes it possible
to fit the observed data to a standard momentum PS
over the wide scale that includes quasi-linear and non-linear scales. 

In section (2.3) we have shown that the radial component 
of the momentum field observed at different locations in the simulation
cube does accurately give the PS of the momentum field 
when the full data (all particles) in the simulation cube is used. 
The situation is different for mock survey data in several aspects. 
First, each mock survey covers only a very small part 
of the whole simulation 
($\sim$ 0.1\% for the MAT survey sample below) 
and the shape of the survey region is conical. 
Second, sampling of the momentum field tracers is sparse (1.4\% at the far
edge $d=50 h^{-1}$ Mpc for the MAT) and not uniform. Third,
the positions of `galaxies' and their peculiar velocities contain
significant and systematic errors.  
The effects of these problems on the PS estimation should be investigated.

For this purpose we have chosen the MAT survey sample 
(Mathewson, Ford \& Buchhorn 1992; Willick et al. 1996). 
The MAP sample is defined by DEC $> -17.5^\circ$ and the galactic latitude
$|b_{G}| > 11^\circ$. We limit the sample at $d< 50 h^{-1}$ Mpc.
Within the survey boundaries there are 814 MAT galaxies 
with diameters $\ge 1.7'$.
We have made mock surveys in the OCDM20$a$ at 100
random locations in the simulation with 
the selection criteria on `galaxies' obtained from the MAT sample.
The radial selection function due to the 
diameter limit of the MAT sample is derived from the redshift
distribution of galaxies. 
We mimic the observational errors in the distance measurement 
by assigning 16\% error in $\log d$ of randomly selected galaxies, 
correct for the IHM bias, and then the sample is cut at 
$r_{\rm max}=50 h^{-1}$ Mpc.
We do not find groups in the mock survey samples
to reduce the distance errors, but instead have adopted the lower 
estimate of the distance error quoted by observers (Mathewson, 
Ford \& Buchhorn 1992; Willick et al. 1996).

The PS of the momentum 
fluctuation is measured from equations (10) and (11), and
that of the density field is also calculated from those equations without
the velocity term.
Figure 4 shows the median values (squares) and 68\% limits  of 
$P_p$ (i.e. $3 P_{p_r}$)
 and $P_{\delta}$ measured from the 100 mock surveys 
(Data are correlated over approximately three neighboring points). 
The solid lines are the true power spectra. 
Deviations from the true values reflect imperfectness and 
finiteness of the mock survey data. 
The dampings of $P_p$ and $P_{\delta}$ at large scales (small $k$) 
are due to finiteness of the sample. The damping of $P_{\delta}$ 
at high $k$ is due to the smoothing effect of the distance error.
The rise of $P_p$ at high $k$ is due to the combined effects 
of distance and peculiar velocity errors.

We use their ratio to the true power spectra 
as the correction factors when we measure the PS for the observed
MAT sample. This method of correction for the systematic effects 
in the observed PS by using mock surveys has been developed by Vogeley
et al. (1992) and Park et al. (1994). 
The correction factor compares with that
shown in Figure 1 of Kolatt \& Dekel (1997). The correction
factor to $P_p$ near $k=0.07 h$ Mpc$^{-1}$ is about 1.4
while it is nearly 7 in their POTENT method. 
It implies that the information in the observed velocity field is 
heavily washed out due to the large smoothing in the POTENT method
compared to our method.

We note in Figure 4 that the agreement with the true PS is good upto the
scale $k\sim 0.05 h^{-1}$ Mpc for $P_p$, and upto $k\sim 0.07 h^{-1}$
Mpc for $P_{\delta}$. As mentioned above, at larger scales the estimated 
power spectra
fall below the true ones because the mock samples lack the large
scale fluctuations. The density PS falls steeply at small scales
because of the large distance errors we put in.
However, the momentum PS is less affected by the distance errors,
which allows us to measure the momentum PS accurately
over fairly wide scales.

Another fact we have observed from this experiment is that there
is a fair amount of correlation between the estimated 
$P_p$ and $P_{\delta}$
even though the sample volume is small. 
Figure 5 is the $P_{p}$ versus $P_{\delta}$ at each
$k$ measured from 100 mock surveys and shows they are correlated
when derived from the same sample. The correlation between 
the density and momentum power spectra makes their ratios less
uncertain, and makes it possible to estimate the $\beta$ 
parameter more accurately.

\section{Application to Observed Data}

\subsection{The MAT sample in the Mark III catalog}

In the previous section we have demonstrated that the momentum
PS can be rather accurately measured from existing data.
To apply our method to a real data we have adopted the MAT 
peculiar velocity sample (Mathewson, Ford \& Buchhorn 1992)
in the Mark III catalog (Willick et al. 1996).
We use the distance data from the forward TF relation corrected
for the IHM.
The original MAT sample contains 1355 spiral galaxies. Applying
the angular position and diameter limits to the sample, we
have 1069 galaxies left which are the black dots in Figure 6$a$.
The radial selection function is calculated from this subset.
When the selection function is calculated, the Einstein-de Sitter 
universe is assumed to calculate diameters of galaxies. 
Application of the distance limit of 50
$h^{-1}$ Mpc leaves us with 814 galaxies which is the final sample.
The MAT sample extends quite close to the Galactic plane, and
shows a large difference in the galaxy number density in the 
north and south. Figure 6$b$ shows the number density variation
as a function of the galactic latitude. To account for this
variation we adopted
an angular selection function
$$\phi_G = \phi_o {\rm dex} [0.05(1-{\rm csc}|b_G|)], \eqno (17)$$
where $\phi_o= 300$ for galaxies
in the south and 550 in the north (solid curves in Figure 6$b$).
The resulting momentum and density power spectra are nearly independent
of this angular selection function except for those at the largest scale
$k=0.025 h$ Mpc$^{-1}$. And the estimated $\beta$ parameter is even 
less dependent on $\phi_G$ because momentum and density power spectra
are affected by it in a similar way.
A better treatment of the galactic extinction is to
use the dust maps of Schlegel et al. (1998) in conjunction with the
prescription described in Hudson (1993) which gives
the change in the diameter for a given amount of extinction.

\subsection{Observed momentum power spectrum}

Figure 7 shows the power spectra (black dots) of the galaxy momentum and 
density fields for the MAT sample. The density PS is measured in the
true distance space rather than in the redshift space.
The 68\% uncertainty limit at each wavenumber is estimated 
from 100 mock surveys in the OCDM20$a$ model.
For a comparison, the velocity
PS measured by Kolatt \& Dekel (1997) from the Mark III data are
plotted as stars at three wave numbers where their PS are statistically
meaningful.  We have shown above that the velocity field
is close to linear at scales $k<0.07 h$ Mpc$^{-1}$ in all
models we have considered. At these scales the momentum PS should
be equal to the velocity PS. In fact, Kolatt \& Dekel's velocity PS
nicely match with our momentum PS at the first two wave numbers.
But at $k=0.172 h$ Mpc$^{-1}$ their velocity PS estimate falls below
our momentum PS. 

In section (3.3) the shape of the momentum PS at $k>0.06 h$ Mpc$^{-1}$ 
is shown to be almost identical for models with different
cosmological parameters and bias factors when the normalization of
the model is reasonable. Using this property of the momentum PS,
we re-estimate the uncertainty limit of the momentum PS. 
The PS of mock surveys are first scaled so that they give the MAT PS 
on average. And then their variance in amplitude is measured over
wavenumbers from $k=0.049 \sim 0.20 h$ Mpc$^{-1}$ by fitting the MAT
PS to the PS of each mock survey.
We estimate from the variance that the amplitude of the 
momentum PS of the MAT sample has 
the 68\% confidence limit of $(1.445, 0.715)$ in factor.
This gives the 68\% ranges of the momentum PS at
wavenumbers 0.049 and 0.074 $h$ Mpc$^{-1}$ of 
$2.67^{+1.18}_{-0.76} \times 10^{10}$
and $8.31^{+3.69}_{-2.37} \times 10^{9}$ km$^2$ sec$^{-2} (h^{-1}
{\rm Mpc})^3$
which are significantly more accurate
than those shown in Figure 7 estimated from the fluctuation  
at each wavenumber.

\subsection{Estimating the $\beta$ parameter}

Calculating both density and momentum power spectra from a given sample
is important because we can measure the $\beta$ 
parameter self-consistently.
The galaxy density PS measured from the MAT sample is also shown 
as filled circles in Figure 7. This is compared with the
galaxy power spectra measured for the CfA redshift survey sample
by Park et al. (1994). The triangles are from the volume-limited
CfA sample with depth of 130 $h^{-1}$ Mpc and the squares from 
the 101 $h^{-1}$ Mpc deep sample. These redshift power spectra
of the CfA galaxy distribution is in good agreement with the 
true space PS from the MAT sample.

Once the density and momentum power spectra are measured from
a sample, one can calculate the $\beta_O = \Omega^{0.6}/b_O$ parameter
from their ratio, where the subscript stands for optical galaxies.
Since these power spectra are nearly in the
linear regime at scales $k\le 0.07 h$ Mpc$^{-1}$, we use the values
at two wavenumbers $k=$ 0.049 and 0.074 $h$ Mpc$^{-1}$ to estimate 
$\beta_O$ from 
the linear relation $\beta^2 = k^2 P_p(k)/H^2 P_{\delta}(k)$.
The $\beta_O$ averaged over the two wavenumbers
is $0.51^{+0.13}_{-0.08}$ where the 68\%
confidence limits are again estimated from the 100 mock survey
samples in the OCDM20$a$ model.
This corresponds to the cosmological density parameter
$\Omega = 0.33^{+0.15}_{-0.09} \; b_O^{5/3}$.
In Figure 8 the MAT momentum PS is multiplied by 
$(k/H\beta_O)^2$ with $\beta_O=0.514$ (open circles) and 1 (squares).
The accuracy of this measured $\beta$ parameter is high
even though the MAT sample is not large. This is because we are able
to accurately measure the amplitude of the velocity PS 
by using the momentum PS over a wide range of wavenumber space, 
and because the estimation of $\beta$ is made using
the density PS measured from the same sample which tends to 
statistically fluctuate in a way similar to the momentum PS.

The amplitude of matter fluctuation is often represented by
the quantity $\sigma_8 \Omega^{0.6} = \sigma_{g,8} \beta$.
An integral over the galaxy density PS from the MAP yields
$\sigma_{g,8}=1.08\pm 0.30$ where the uncertainty limit is
from the mock surveys. This gives $\sigma_8 \Omega^{0.6} = 0.56
\pm 0.21$.

\section{Discussions}

We have found that the radial peculiar velocity/momentum field can be
considered as a scalar field whose PS is very close to
1/3 of that of the three-dimensional isotropic velocity/momentum field.
Considering the fact that the observed radial velocity is not sampled
smoothly over the space but traced by galaxies, we have adopted 
to use the momentum field which is defined as the peculiar 
velocity weighted by number of galaxies.
This momentum field is equal to the velocity field in the
linear regime. 
We have demonstrated that the momentum field PS measured from
the currently available peculiar
velocity data can give us reasonably accurate estimates
of the matter PS.
We have then calculated the momentum PS for the
MAT peculiar velocity sample of spiral galaxies, and measured
the amplitude of matter PS and the parameter
$\beta_O= 0.51^{+0.13}_{-0.08}$ or the cosmological density parameter
$\Omega = 0.33^{+0.15}_{-0.09} \; b_O^{5/3}$ where $b_O$ is the bias
factor for the MAT galaxies.

There have been several recent measurements of the amplitude
of the matter PS and the $\beta$ parameter. 
The POTENT method has been extensively applied to analyze observational
data (Dekel et al. 1999).
In this method the radial peculiar velocities of galaxies are 
smoothed over a large scale (typically $\ge 12 h^{-1}$ Mpc). 
Then the velocity and density fields are reconstructed using the quasi-linear  
solution of the continuity equation and assuming a potential flow. 
A drawback of this method is that the heavy smoothing makes
the resulting data informative only over a narrow spatial range.
For example, the Mark III catalog has been
used to probe the velocity field out to about $60 h^{-1}$ Mpc  
by Kolatt \& Dekel (1997). If such data is smoothed by a 12 $h^{-1}$ Mpc 
Gaussian as Kolatt \& Dekel did, the remaining usable dynamic range
is roughly between $30\sim 100 h^{-1}$ Mpc.
In the POTENT analysis the heavy smoothing is necessary to reduce 
the small-scale non-linearities and the random distance errors, 
but more importantly to interpolate the missing velocities within voids. 
The smoothing is also not a simple procedure since the radial peculiar 
velocity vectors merge towards the observer. 

Kolatt \& Dekel have presented $\beta_O = 0.80\pm 0.10$ for Mark III galaxies
by using the PS of the APM galaxies (Tadros \& Efstathiou 1995), and 
$0.77\pm 0.11$ using the CfA2+SSRS2 galaxy PS (Park et al. 1994). 
Our momentum PS is consistent with their velocity PS in the linear
regime. Despite this fact, their estimate of the $\beta$
parameter is much higher than ours. The major difference in the
estimation of $\beta$ is that they have simply adopted the density PS
calculated from various redshift surveys while we have calculated it
from the same sample where the momentum PS has been measured.
An earlier POTENT application to the Mark III data is Hudson et al.
(1995) who have obtained $\beta_O=0.74\pm 0.13$, which is also
higher than our estimate.
Freudling et al. (1999) has applied the maximum-likelihood method 
to the SFI sample, and found a value of $\beta=0.82\pm 0.12$, 
which is close to the POTENT results.

The velocity CF statistic can be directly calculated from observed 
discrete data without smoothing.  Borgani et al. (2000) has calculated 
the velocity CF from the SFI sample to find $\sigma_8 \Omega^{0.6}
=0.3 \pm 0.1 (\Gamma/0.2)^{0.5}$ where $\Gamma$ is the shape parameter
of the CDM models. 
The observed peculiar velocities and absolute distances contain 
large errors. Since the CF at a certain scale is determined by the power 
at all other scales, the large small-scale random 
errors can affect the velocity CF at all scales.
As in the case of galaxy distribution studies, the velocity PS
gives more accurate measure of clustering at large scales
and is more directly comparable to cosmological models compared to the CF.
Another low $\beta_O$ estimation has been reported by Blakeslee et al. 
(1999) who have found $\beta_O=0.26\pm 0.08$ using a peculiar
velocities in a surface brightness fluctuation survey of galaxy 
distances and the optical redshift survey (ORS; Santiago et al. 1995).
A similarly low value $\beta_O=0.3\pm 0.1$ has been found by Riess et al. 
(1997) who have used the velocity field from Type-Ia supernovae
and that predicted from the ORS galaxies.
These results are inconsistent with those from the POTENT or the
maximum-likelihood methods considering their quoted error bars.
Despite their small uncertainty limits the recent studies altogether
allow $\beta_O$ to be in the range $0.2\sim 1$ or roughly $\Omega=0.05\sim 1
\; b_O^{5/3}$, which are not very interesting constraints.

Our result is consistent with the result $\beta_O =0.50\pm 0.06$
of Hudson (1994) who has compared the observed peculiar velocity
with the predicted velocity field from galaxy density field 
using the gravitational instability linear theory. 
A power spectrum analysis of Gramann (1998) has also yielded
$\beta_O = 0.5 - 0.6$ for the Stromlo-APM redshift survey.

Interestingly, a good agreement with other studies in $\beta$ parameter 
estimation is found when the IRAS galaxies are used 
to infer the gravity field, even though
$\beta_I/\beta_O$ could be as large as 1.6 (Blakeslee et al. 1999). 
Davis et al. (1996) has compared the IRAS gravity field with
the velocity field of Mark III spirals calculated from the orthogonal
mode-expansion method, and determined $\beta_I$ in
the most likely range of $0.4\sim 0.6$.
On the other hand,
da Costa et al. (1998) has compared the velocity field directly
measured from the SFI spiral galaxy survey with that derived from
the IRAS 1.2-Jy galaxy redshift survey, and obtained 
$\beta_I=0.6\pm 0.1$. When they move the outer sample boundary
from 6000 to 4000 km s$^{-1}$, they found a smaller $\beta_I\approx 0.5$.
More recently, Nusser et al. (2000) has used the ENEAR peculiar
velocity sample and the IRAS {\it PSCz} redshift survey sample to
obtain $\beta_I=0.5\pm 0.1$. Somewhat lower estimates of $\beta_I$ are
found by Blakeslee et al. (1999) and Riess et al. (1997), who have reported
$\beta_I=0.42^{+0.10}_{-0.06}$ and $0.40\pm 0.15$, respectively.

It is to be noted that our estimate for the mass density
parameter $\Omega\sim 0.33$ is in a good agreement with the constraint
from the Cosmic Microwave Background data (cf. Tegmark and Zaldarriaga
2000) if the bias factor has a reasonable value around one.
We plan to apply our analysis method to more recent wide angle samples 
like the SFI and the ENEAR samples. With these larger survey samples
we hope to be able to find the amplitude of mass fluctuation more accurately,
and also to study the difference in the velocity/momentum fields 
traced by galaxies with different morphology.

\acknowledgments

This work was supported by the KOSEF grant (1999-2-113-001-5).
The author would like to thank Canadian Institute for
Theoretical Astrophysics for the hospitality during this work, 
David H. Weinberg and Michael J. Hudson for helpful comments, and 
the anonymous referee for giving many valuable comments and suggestions.

\clearpage


\begin{figure}
\plotone{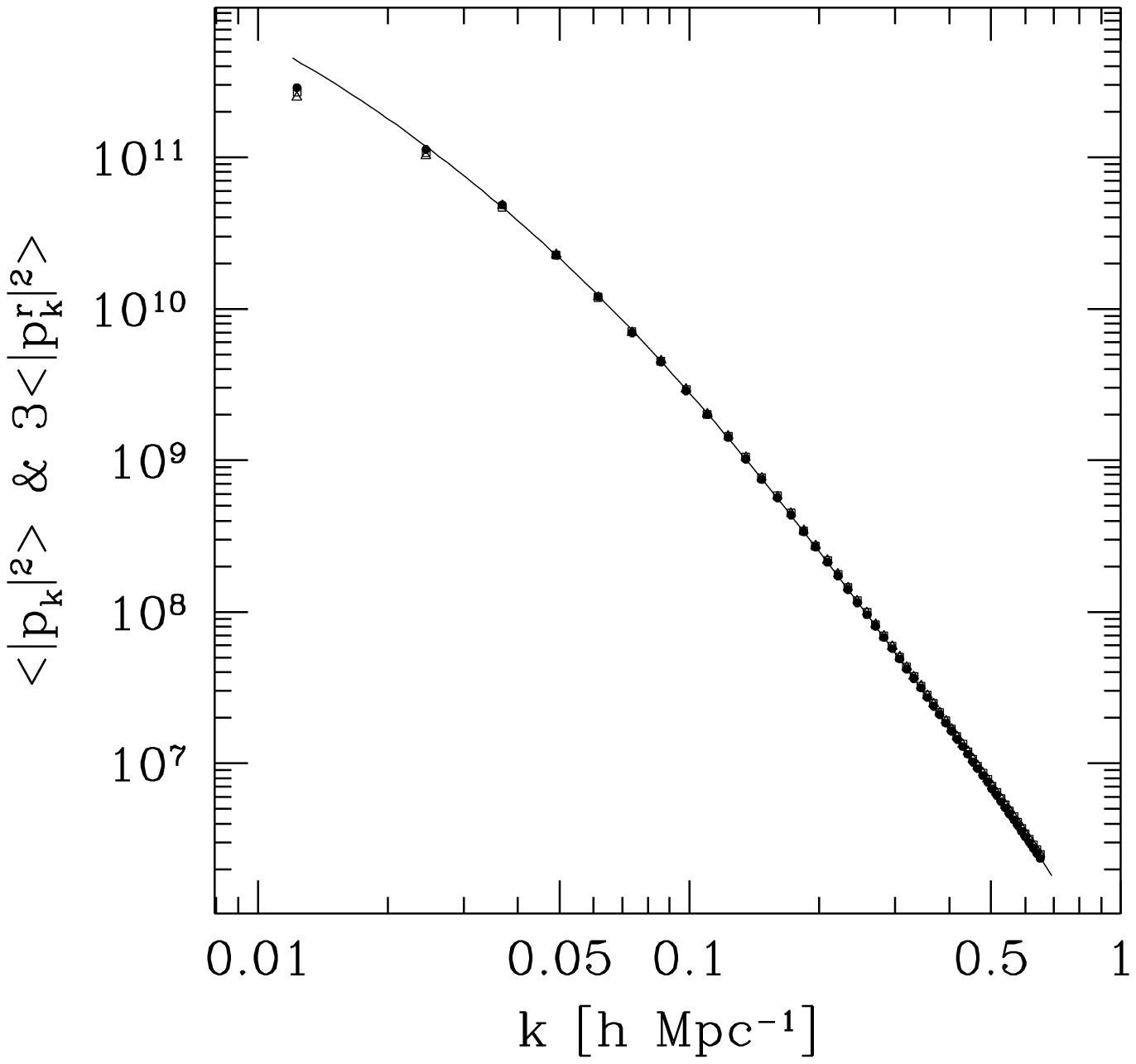}
\caption{The power spectrum of the three-dimensional momentum field 
(filled circles), and three times the power spectra of the radial
component of the momentum field observed at the center (triangles)
and at a corner (squares) of the simulation cube. Twenty linear realizations
of an open CDM model with $\Omega h=0.2$ in a 512 $h^{-1}$ Mpc box
are used to obtain the average power spectra.
The solid line is the theoretical power spectrum of the model.}
\end{figure}

\begin{figure}
\plotone{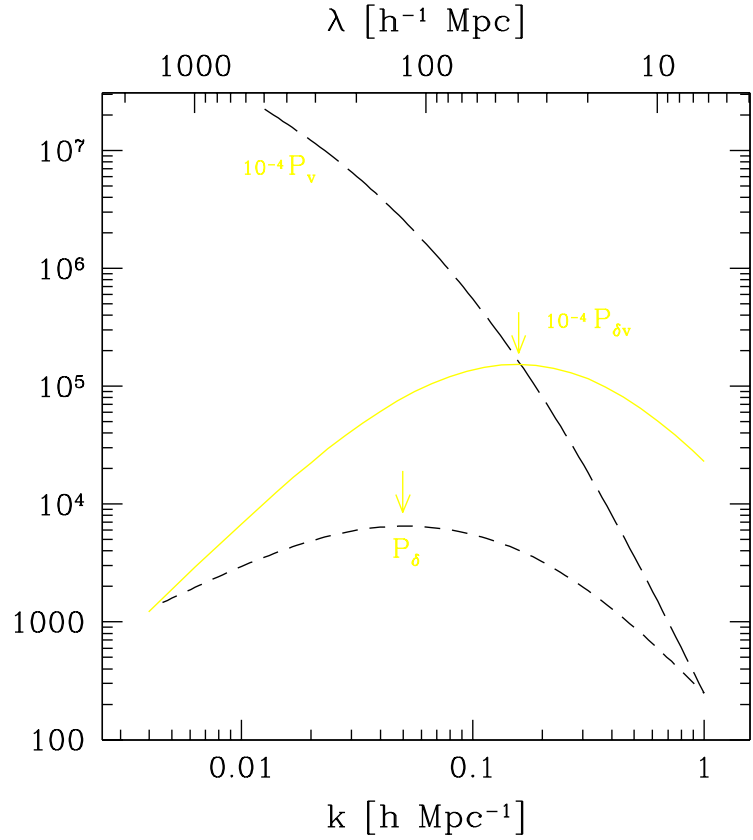}
\caption{The linear density and velocity power spectra $P_{\delta}$ and
$P_v$ in the SCDM cosmogony.  The corresponding power spectrum of
the $\delta {\bf v}$ field calculated from equation (14)
is also shown (solid curve).
Arrows indicate locations of the maxima of the power spectra.
}
\end{figure}

\begin{figure}
\plotone{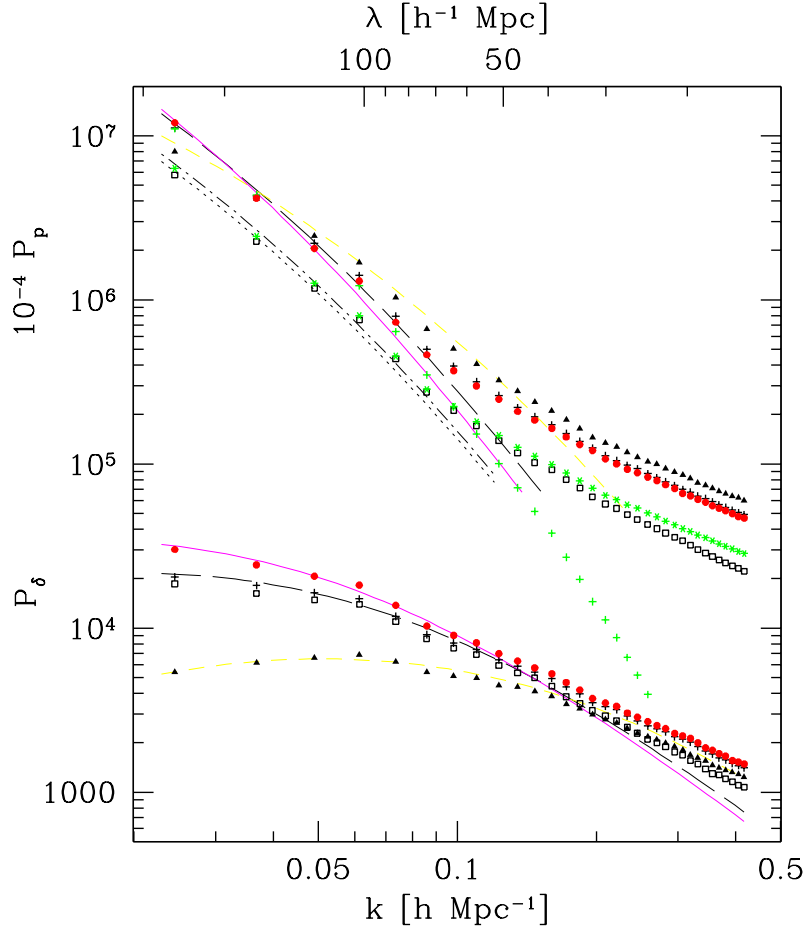}
\caption{True power spectra of the momentum and density fields of the
simulated universe models calculated from exact positions and
peculiar velocities of all particles or `galaxies' in the simulations. 
Models shown are the SCDM (triangle symbols, 
short dashed curves), OCDM20a (upper crosses, long-dashed), 
OCDM15 (filled circles, solid) models with $b=1$ and $h=0.5$. 
Symbols indicates the power spectra calculated from simulations
of $256^3$ CDM particles in a 512 $h^{-1}$ Mpc box.
Curves are the linear power spectra. 
The OCDM20b (squares, dotted) has $b=1.4$, and the OCDM20c 
(stars, dot-dashed) has $b=1$, $\Omega = 0.15$ and $h=0.8$. 
The linear density power spectra
of OCDM20b and OCDM20c models are the same as that of OCDM20a.
The velocity power spectrum $P_v$ has been also shown for the OCDM20a 
simulation (lower crosses diverging from $P_p$).
}
\end{figure}

\begin{figure}
\plotone{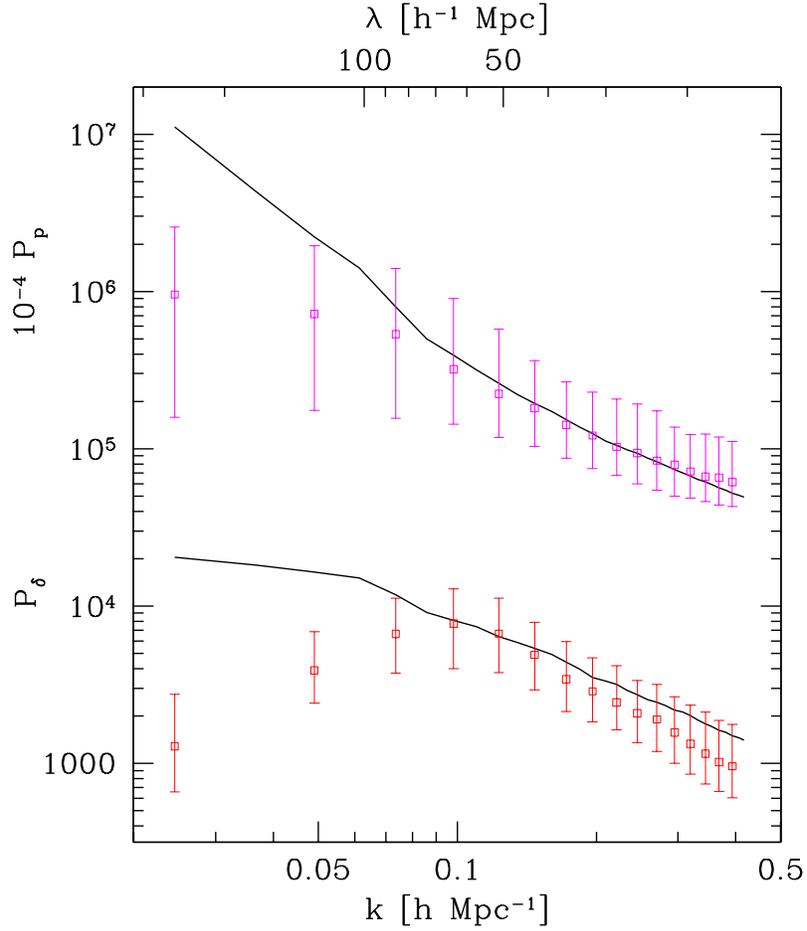}
\caption{The median values (squares) and 68\% confidence limits of the
momentum and density fields obtained from 100 mock MAT surveys in the 
OCDM20a model. 
The solid curves are the true power spectra.
Power spectra are correlated approximately over three neighboring points.
Each mock survey has the geometry of the MAT survey and is limited
to distance of 50 $h^{-1}$ Mpc.
The `galaxies' in the mock surveys are randomly perturbed in 
${\rm log} d$ by 16\% where $d$ is the true distance.
}
\end{figure}

\begin{figure}
\plotone{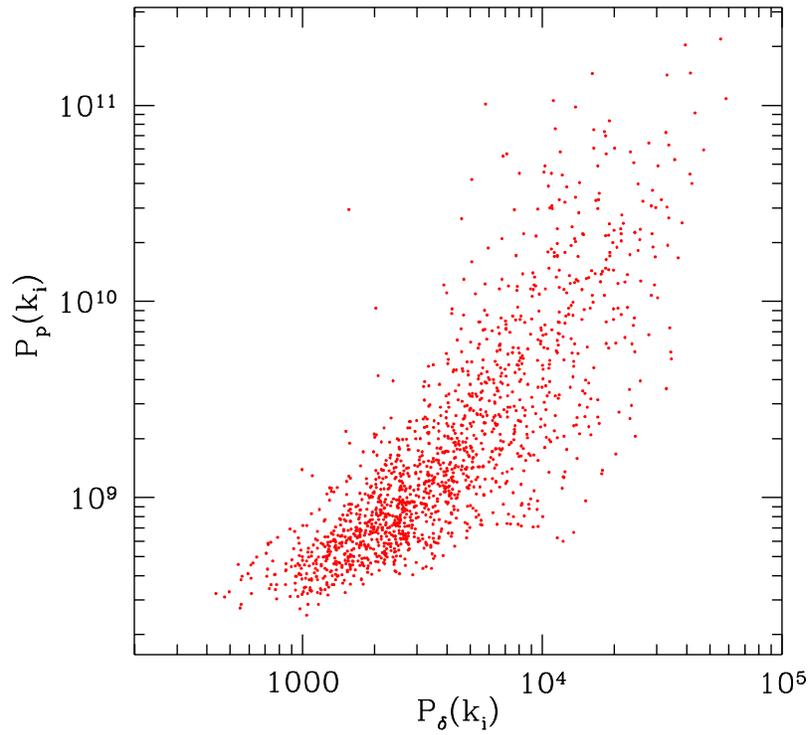}
\caption{Correlation of the momentum power spectrum with the density
power spectrum at each wavenumber. A pair of power spectra 
$P_{\delta}$ and $P_p$ are calculated 
from each of the 100 mock MAT surveys in the OCDM20a model.
}
\end{figure}

\begin{figure}
\plotone{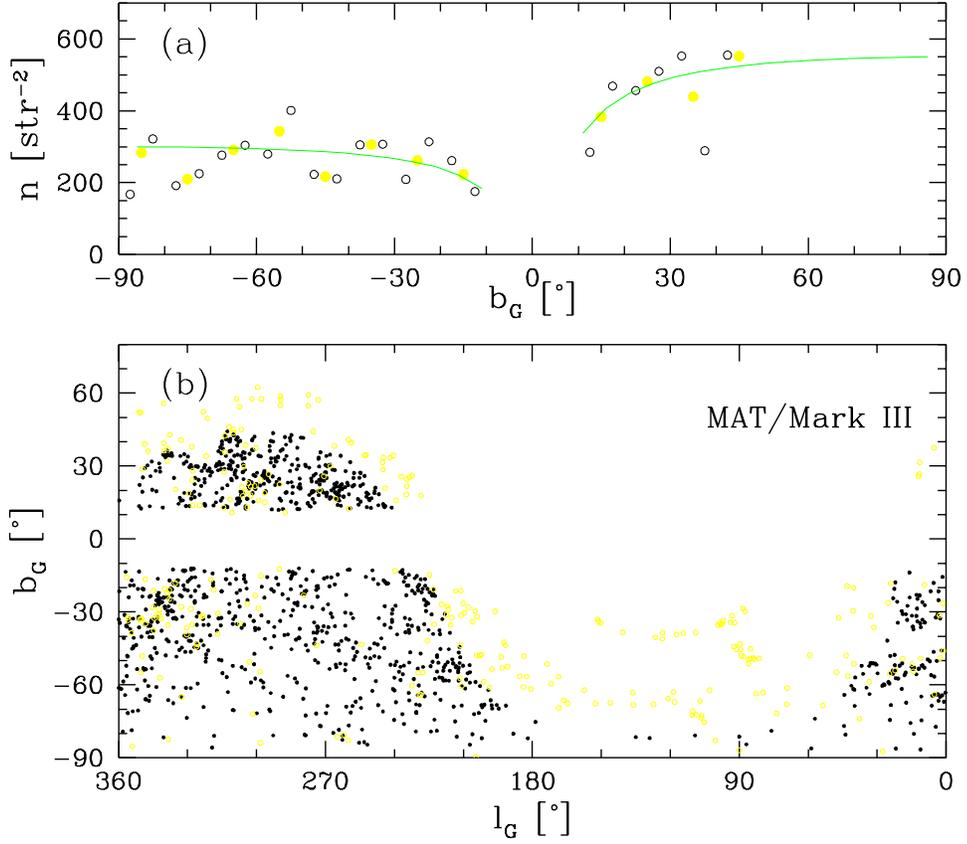}
\caption{(a) Mean number density of galaxies in the MAT catalog
as a function of the Galactic latitude. Averages are taken over
$5^\circ$ (open circles) and $10^\circ$ (filled circles).
The solid curves are the functions $\phi_G = \phi_o {\rm dex}[0.05(1-
{\rm csc}|b_G|)]$ where $\phi_o=300$ in the south and 550 in the north.
(b) Distribution of MAT galaxies on the sky in the Galactic coordinate. 
The filled dots are the 1069 galaxies satisfying the angular 
and diameter limits.}
\end{figure}

\begin{figure}
\plotone{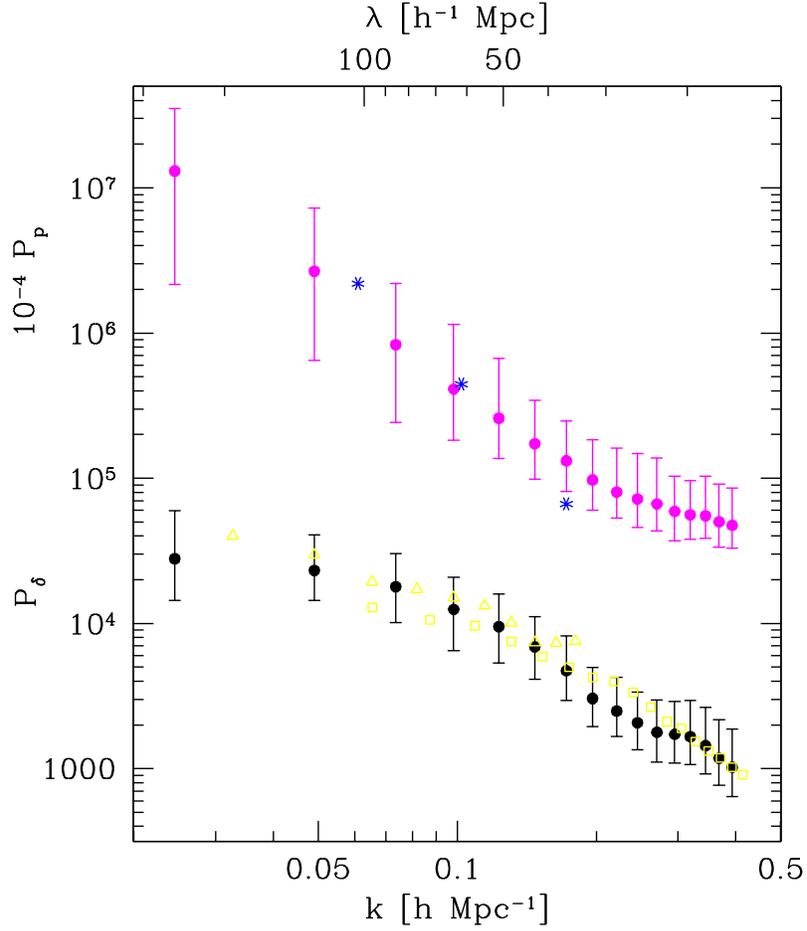}
\caption{The momentum (upper filled circles) and density (lower) power spectra
measured from the MAT sample. The 68\% uncertainty limits are derived
from 100 mock MAT surveys in the OCDM20a model. Stars are the velocity
power spectrum measured from the Mark III catalog by Kolatt \& Dekel (1997).
Triangles and squares are the galaxy density power spectra in the redshift space 
calculated by Park et al. (1994) for the 130 and 
101 $h^{-1}$ Mpc deep sub-samples of the CfA survey, respectively.
}
\end{figure}

\begin{figure}
\plotone{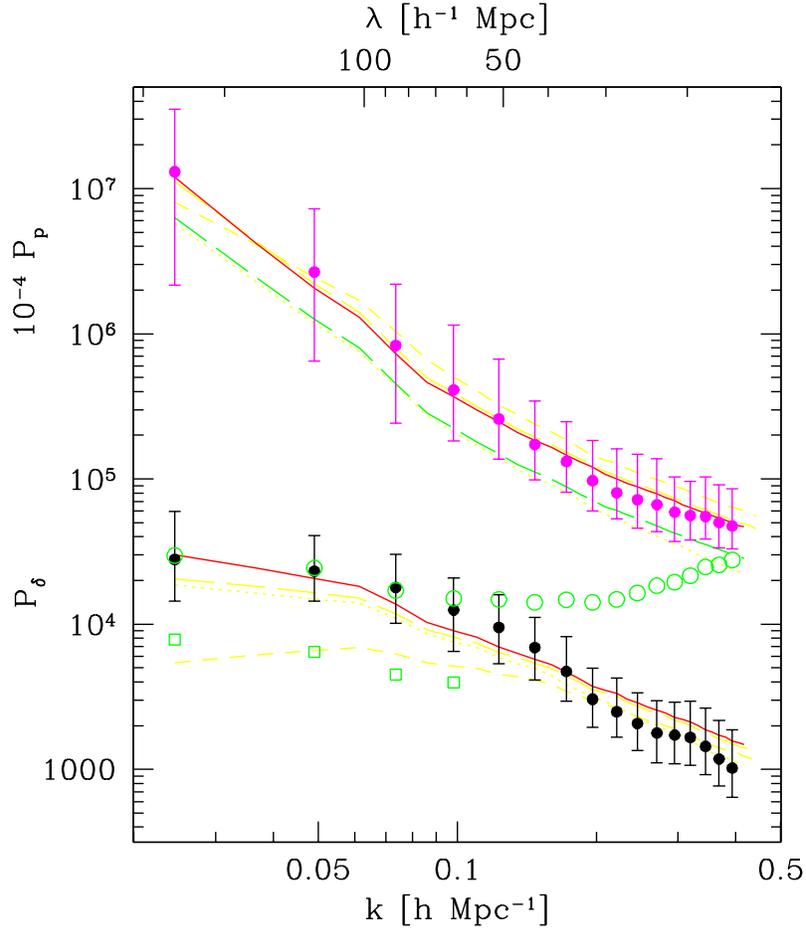}
\caption{Momentum and density power spectra of the simulated universe models
and of the galaxies in the MAT sample (filled circles with 68\% error bars).
Models compared are the SCDM (dashed curves), OCDM20a (long-dashed), OCDM15 (solid),
OCDM20b (dotted), and OCDM20c (lower long-dashed $P_p$).
The open circles are $(k/H\beta_O)^2 P_p(k)$ with $\beta_O=0.51$ for the MAT
sample. The open squares are the case with $\beta_O=1.0$.
}
\end{figure}

\end{document}